\documentclass[usenatbib,useAMS]{mn2e}

\newcommand{\mc}[3]{\multicolumn{#1}{#2}{#3}}
\newcommand{\pmean}{\langle p \rangle}
\newcommand{\mcb}[1]{\multicolumn{1}{c}{#1}}

\usepackage{color}
\usepackage{graphicx}
\usepackage{amsmath}
\usepackage{amssymb}
\usepackage{perpage} 
\usepackage{pdflscape}
\MakePerPage{footnote} 

\title[RoboPol: Blazar optical polarization survey]{The RoboPol optical polarization survey of gamma-ray--loud blazars}
\author[V. Pavlidou et al.]
{V. Pavlidou$^{1,2}$\thanks{Contact authors' e-mail addresses:
pavlidou@physics.uoc.gr (VP); eangelakis@mpifr-bonn.mpg.de (EA)}, E. Angelakis$^{3\star}$, I. Myserlis$^{3}$, D. Blinov$^{2,7}$, O.\,G. King$^{4}$, I. Papadakis$^{2,1}$, 
\newauthor
K. Tassis$^{2,1}$,T. Hovatta$^{4,8}$, B. Pazderska$^5$,   E. Paleologou$^{2}$, M. Balokovi\'{c}$^4$, R. Feiler$^5$, 
\newauthor
L. Fuhrmann$^3$,   P. Khodade$^6$, A. Kus$^5$, N. Kylafis$^{2,1}$,  D. Modi$^{6}$, G. Panopoulou$^2$,  
\newauthor
 I. Papamastorakis$^{2,1}$,  E. Pazderski$^5$, T.\,J. Pearson$^{4}$, C. Rajarshi$^6$, A.  Ramaprakash$^6$,
\newauthor A.\,C.\,S. Readhead$^{4}$, P. Reig$^{1,2}$, J.\,A. Zensus$^3$\\
$^{1}$Foundation for Research and Technology - Hellas, IESL, Voutes, 7110 Heraklion, Greece\\
$^{2}$Department of Physics and Institute for Plasma Physics, University of Crete, 71003, Heraklion, Greece\\
$^{3}$Max-Planck-Institut f\"{u}r Radioastronomie, Auf dem H\"{u}gel
69, 53121 Bonn, Germany\\
$^{4}$Cahill Center for Astronomy and Astrophysics, California Institute of Technology, 1200 E California Blvd, MC 249-17,\\Pasadena CA, 91125, USA\\
$^5$Toru\'{n} Centre for Astronomy, Nicolaus Copernicus University, Faculty of Physics, Astronomy and Informatics,\\
Grudziadzka 5, 87-100 Toru\'{n}, Poland\\
$^6$Inter-University Centre for Astronomy and Astrophysics, Post Bag
4, Ganeshkhind, Pune - 411 007, India\\
$^7$Astronomical Institute, St. Petersburg State
University,Universitetsky pr. 28, Petrodvoretz, 198504 St. Petersburg,
Russia\\
$^8$ Aalto University Mets\"ahovi Radio Observatory, Mets\"ahovintie 114, 02540 Kylm\"al\"a, Finland
}
\begin{document}



\maketitle

\label{firstpage}

\begin{abstract} We present first results from RoboPol, a novel-design optical polarimeter
  operating at the Skinakas Observatory in Crete. The data, taken during the May -- June
  2013 commissioning of the instrument, constitute a single-epoch linear polarization
  survey of a sample of gamma-ray--loud blazars, defined according to unbiased and
  objective selection criteria, easily reproducible in simulations, as well as a
  comparison sample of, otherwise similar, gamma-ray--quiet blazars. As such, the results
  of this survey are appropriate for both phenomenological population studies and for
  tests of theoretical population models. We have measured polarization fractions as low
  as $0.015$ down to $R$ magnitude of 17 and as low as $0.035$ down to 18 magnitude.  The
  hypothesis that the polarization fractions of gamma-ray--loud and gamma-ray--quiet
  blazars are drawn from the same distribution is rejected at the $3\sigma$ level. We
  therefore conclude that gamma-ray--loud and gamma-ray--quiet sources have different
  optical polarization properties.  This is the first time this statistical difference is demonstrated in optical wavelengths. The polarization fraction distributions of both
  samples are well-described by exponential distributions with averages of $\pmean =6.4
  ^{+0.9}_{-0.8}\times 10^{-2}$ for gamma-ray--loud blazars, and $\pmean =3.2
  ^{+2.0}_{-1.1}\times 10^{-2}$ for gamma-ray--quiet blazars. The most probable value for
  the difference of the means is $3.4^{+1.5}_{-2.0}\times 10^{-2}$. The distribution of
  polarization angles is statistically consistent with being uniform. \end{abstract}


\section{Introduction}\label{intro}

Blazars, which include BL Lac objects and Flat Spectrum Radio Quasars (FSRQs), represent
the class of gamma-ray emitters with the largest fraction of members associated with known
objects \citep{2012ApJS..199...31N}. They are active galactic nuclei with their jets
closely aligned to our line of sight \citep{RogerArieh}. Their emission is thus both
beamed and boosted through relativistic effects, so that a large range of observed
properties can result from even small variations in their physical conditions and
orientation. As a result, the physics of jet launching and confinement, particle
acceleration, emission, and variability, remain unclear, despite decades of intense
theoretical and observational studies.

Blazars are broadband emitters exhibiting spectral energy distributions ranging from cm
radio wavelengths to the highest gamma-ray energies \citep[e.g.][]{2012A&A...541A.160G}
with a characteristic ``double-humped'' appearance. While the mechanism of their
high-energy (X-ray to gamma-ray) emission remains debatable, it is well established that
lower-energy jet emission is due to synchrotron emission from
relativistic electrons. Linear polarization is one property
characteristic of the low-energy emission.

Polarization measurements of blazar synchrotron emission can be challenging, yet
remarkably valuable. They probe parts of the radiating magnetised plasma where the
magnetic field shows some degree of uniformity quantified by $\frac{B_0}{B}$ where $B_0$
is a homogeneous field and $B$ is the total field \cite[e.g.][]{1972Ap&SS..19...25S}.
The polarized radiation then carries information about the structure of the magnetic field
in the location of the emission (strength, topology and uniformity). Temporal changes in
the degree and direction of polarization can help us pinpoint the location of the emitting
region and the spatiotemporal evolution of flaring events within the jet.

Of particular interest are rotations of the polarization angle in optical wavelengths
during gamma-ray flares, instances of which have been observed through polarimetric
observations concurrent with monitoring at GeV and TeV energies, with {\it Fermi}-LAT
\citep{atwood} and MAGIC \citep{magic} respectively \citep[e.g.,][]{FermiNature,
  MarscherNature}. If such rotations were proven to be associated with the outbursting
events of the gamma-ray emission, then the optopolarimetric evolution of the flare could
be used to extract information about the location and evolution of the gamma-ray emission
region.

Such events have stimulated intense interest in the polarimetric monitoring of gamma-ray
blazars (\citealp[e.g.,][]{SPbSU,Arizona,Kanata}). These efforts have been focusing more
on ``hand-picked'' sources and less on statistically well-defined samples aiming at
maximising the chance of correlating events. Consequently, although they have resulted in
the collection of invaluable optopolarimetric datasets for a significant number of
blazars, they are not designed for rigorous statistical studies of the blazar population;
the most obvious one being the investigation of whether the observed events are indeed
statistically correlated with gamma-ray flares, or are the result of chance
coincidence. The RoboPol program has been designed to bridge this gap.

The purpose of this paper is two-fold. Firstly, we aim to present the results of a survey that
RoboPol conducted in June 2013, which is the first single-epoch optopolarimetric survey of
an unbiased sample of gamma-ray--loud blazars. As such, it is appropriate for
statistical phenomenological population studies and for testing blazar population
models. Secondly, we wish to alert the community to our optopolarimetric monitoring program and to
encourage complementary observations during the Skinakas winter shutdown of December --
March.

After a brief introduction to the RoboPol monitoring program in Section \ref{sec:program},
the selection criteria for the June 2013 survey sample and the July -- November 2013
monitoring sample are reviewed in Section \ref{sample}. The results from the June 2013
survey are presented in Section \ref{observations}, where the optical polarization
properties of the survey sample and possible differences between gamma-ray--loud and
gamma-ray--quiet blazars are also discussed.  We summarise our findings in Section
\ref{discussion}.

\section{The RoboPol optopolarimetric monitoring program}
\label{sec:program}
The RoboPol program has been designed with two guiding principles in mind:
\begin{enumerate}
\item to provide datasets ideally suited for rigorous statistical studies;
\item to maximise the potential for the detection of polarization rotation events.
\end{enumerate} 
To satisfy the former requirement, we have selected a large sample of blazars on the basis
of strict, bias-free, objective criteria, which are discussed later in this paper. To
satisfy the latter, we have secured a considerable amount of evenly allocated telescope
time; we have constructed a novel, specially designed polarimeter -- the RoboPol
instrument -- (A. N. Ramaprakash et al.~, in preparation, hereafter ``instrument'' paper);
and we have developed a system of automated telescope operation including data reduction
that allows the implementation of dynamical scheduling (\citealt[][]{PipelinePaper}, hereafter
``pipeline'' paper). The long-term observing strategy of the RoboPol program is the monitoring of
$\sim$ 100 target (gamma-ray--loud) sources and an additional $\sim$15 control (gamma-ray--quiet)
sources with a duty cycle of about 3 nights for non-active sources and several times a
night for sources in an active state.

\subsection{The RoboPol instrument}

The RoboPol instrument (described in the ``instrument'' paper) is a novel-design 4-channel
photopolarimeter. It has no moving parts, other than a filter wheel, and simultaneously
measures both linear fractional Stokes parameters $q=Q/I$ and $u=U/I$. 
This design bypasses the need for multiple exposures with different half-wave plate
positions, thus avoiding unmeasurable errors caused by sky changes between measurements
and imperfect alignment of rotating optical elements.  
The instrument has a $13'\times13'$ field of view, enabling relative photometry using
standard catalog sources and the rapid polarimetric mapping of large sky areas. It is
equipped with standard Johnson-Cousins $R$- and $I$-band band filters from Custom
Scientific. The data presented in this paper are taken with the $R$-band filter. RoboPol
is mounted on the 1.3-m, f/7.7 Ritchey--Cretien telescope at Skinakas Observatory
\citep[1750\,m, $23^\circ53'57''′$E, $35^\circ12'43''$N,][]{2007Ippa....2b..14P} in Crete,
Greece. It was commissioned in May 2013.
\begin{table*}
 \centering
  \caption{\label{sample_select} Selection criteria for the gamma-ray--loud and the
    control sample. A summarising chart is shown in Fig.~\ref{chart}.}
  \begin{tabular}{|p{.3\textwidth}|p{.3\textwidth}|p{.3\textwidth}|} 
     \hline
  Property             &Allowed range for the June survey     &Allowed range for the 2013 monitoring          \\
  \hline
\\
\multicolumn{3}{c}{\bf Gamma-ray--loud sample}\\\\[-1.5ex]
2FGL $F\left(>100\right)$\,MeV           &$> 2\times
10^{-8}{\rm\,cm^{-2}\,s^{-1}}$ & $> 2\times 10^{-8}{\rm\,cm^{-2}\,s^{-1}}$ \\\\[-1.5ex]
2FGL source class                        &agu, bzb, or bzq                   &agu, bzb, or bzq                     \\\\[-1.5ex]
Galactic latitude $|b|$                  &$> 10^\circ$                     & $> 10^\circ$                     \\\\[-1.5ex]
Elevation (Elv) constraints$^1$  &$\mathrm{Elv}\geq 30^\circ$ for at least 30\,min in June &$\mathrm{Elv_{max}}\geq 40^\circ$ for at least 120 consecutive days in the window June -- November including  June\\\\[-1.5ex] 
$R$ magnitude                   & $\leq 18^{2}$ & $\leq 17.5^{3}$ \\\\
\multicolumn{3}{c}{\bf Control sample}\\\\[-1.5ex]
CGRaBS/15\,GHz OVRO monitoring               &included     &included\\\\[-1.5ex]
2FGL                                         &not included &not included \\\\[-1.5ex]
Elevation constraints$^1$                     &None          &$\mathrm{Elv_{max}}\geq 40^\circ$ constantly in the window mid-April -- mid-November\\\\[-1.5ex]
$R$ magnitude                             & $ \leq 18$  & $ \leq 17.5^{2}$ \\\\[-1.5ex]
OVRO 15\,GHz mean flux density               &N/A          &$\geq 0.060$\,Jy \\\\[-1.5ex]
OVRO 15\,GHz intrinsic modulation index, $m$ &$\geq 0.02$  &$\geq 0.05$ \\\\[-1.5ex]
Declination                                  & $\geq 54.8^\circ$ (circumpolar)  &N/A\\\\
\hline
\end{tabular}
\begin{flushleft}                                                                                
$^1$Refers to elevation during Skinakas dark hours\\                                         
$^2$Archival value\\                                                       
$^3$Average value between archival value and measured during preliminary RoboPol Skinakas observations in June 2012 (when applicable)                                             
\end{flushleft}                                                                                  
\end{table*}

\subsection{The first RoboPol observing season}
In June 2013 RoboPol performed an optopolarimetric survey of a sample of gamma-ray--loud
blazars, results from which are presented in this paper. Until November 2013, it was
regularly monitoring (with a cadence of once every few days) an extended sample of
blazars, described in Section~\ref{sample}.  
These sources were monitored until the end of the observing season at Skinakas
(November 2013). The results of this first-season monitoring will be discussed in an
upcoming publication.

\subsection{Multi-band monitoring of the RoboPol sources}

All of our sources (including the control sample) are monitored twice a week at 15\,GHz by
the OVRO 40-m telescope blazar monitoring program \citep{RichardsEtal2011}. 28 of them are
also monitored at 30\,GHz by the Toru\'{n} 32-m telescope
\citep[e.g.][]{2000SPIE.4015..299B,Torun}. Additionally, our sample includes most sources
monitored by the F-GAMMA program \citep{2007AIPC..921..249F,2010arXiv1006.5610A} that are
visible from Skinakas; for these sources, the F-GAMMA program takes multi-band radio data
(total power, linear and circular polarization) approximately once every 1.3 months. By
design, {\it Fermi}-LAT in its sky-scanning mode is continuously providing gamma-ray data
for all of our gamma-ray--loud sources. In this way, our sample has excellent multi-band
coverage. These multiwavelength data will be used in the future to correlate the behaviour of our sample in optical flux and polarization with the properties and variations in other wavebands.

\section{Sample selection criteria}\label{sample}
\subsection{Parent Sample}
We construct a gamma-ray flux-limited ``parent sample'' of gamma-ray--loud blazars from the
second {\it Fermi}-LAT source catalog \citep{2012ApJS..199...31N} using sources tagged
as BL Lac (bzb), FSRQ (bzq), or active galaxy of uncertain type (agu). The
parent sample is created the following way:
\begin{enumerate}
\item for each source, we add up {\it Fermi}-LAT fluxes above 100\,MeV to obtain the
  integrated photon flux $F\left(>100\mathrm{\,MeV}\right)$,
\item we exclude sources with $F\left(>100{\rm\,MeV}\right)$ less than
  or equal to
  $2\times 10^{-8}$\,cm$^{-2}$\,s$^{-1}$, and
\item we exclude sources with galactic latitude $|b|\leq 10^\circ$.
\end{enumerate}
This leaves us with 557 sources in the parent sample. We have verified that the sample is
truly photon-flux-limited since there is no sensitivity dependence on spectral index or
galactic latitude with these cuts. Of these 557 sources, 421 are ever observable from Skinakas: they have at least one night with airmass less than 2, (or, equivalently, elevation higher
than $30^\circ$), for at least one hour, within the dark hours of the May -- November
observing window. Archival optical magnitudes were obtained for all 557 sources in the
parent sample mostly in the $R$-band using the BZCAT \citep{Massaro2009yCat_34950691M},
CGRaBS \citep{CGRaBS}, LQAC 2 \citep{2011yCat..35379099S} and GSC 2.3.2
\citep{2008AJ....136..735L} catalogs \footnote[1]{2 sources were found in $V$-band, 8
  in $B$-band and 2 in $N$-band (0.8\,$\mu$m)}.
 
\subsection{June 2013 Survey Sample}\label{jss}
The June 2013 survey sample was constructed of parent sample sources with a recorded
archival $R$ magnitude less or equal to 18 ($R \leq 18$) which were visible from Skinakas
during dark hours in the month of June 2013 for at least 30\,min at airmass less than
2. The selection criteria for the candidate sources in this sample are summarised in
Table~\ref{sample_select}. This selection resulted in 142 sources potentially observable
in the month of June which constitute a {\em statistically complete} sample. The sources
were observed according to a scheduling algorithm designed to maximise the number of
sources that could be observed in a given time window based on rise and set times,
location of sources on the sky, and resulting slewing time of the telescope. At the end of
the survey, 133 of these sources had been observed. Because the scheduling algorithm was
independent of intrinsic source properties, the resulting set of 133 observed sources is
an {\em unbiased} subsample of the statistically complete sample of the 142 sources
(summary in Fig.~\ref{chart}). The completeness of the sample is $93\,\%$ for sources
brighter than 16 magnitude, $95\,\%$ ($81/85$) for sources brighter than 17 magnitude, and
$94\,\%$ ($133/142$) for sources brighter than 18 magnitude.

To identify sources suitable for inclusion in the ``control'' sample, a number of non-2FGL
CGRaBS \citep[Candidate Gamma-Ray Blazar Survey,][]{CGRaBS} blazars were also observed
during the June survey. CGRaBS was a catalog of likely gamma-ray--loud
sources selected to have similar radio and X-ray properties with then
known gamma-ray--loud blazars. However, because Fermi has a much
improved sensitivity at higher energies than its predecessors,
Fermi-detected blazars include many sources absent from CGRaBS, with
harder gamma-ray--spectra than CGRaBS sources,
especially at lower gamma-ray fluxes. To ensure that non-CGRaBS
sources among our gamma-ray--loud sample do not affect our conclusions
when populations of gamma-ray--loud and gamma-ray--quiet blazars are
found to have significantly different properties, in such cases we
will also be performing comparisons between our ``control'' sample and
that fraction of our gamma-ray--loud sample that is also included in
CGRaBS. 

Candidate sources for these observations were selected according
to the criteria listed in table~\ref{sample_select}. There, the radio variability
amplitude is quantified through the {\em intrinsic modulation index} $m$ as defined by
\cite{RichardsEtal2011}, which measures the flux density standard deviation in units of
the mean flux at the source. For the sources discussed in Section~\ref{observations}, $m$
is reported in table~3. 
The criteria of table~\ref{sample_select} result in
a statistically complete sample of 25 in principle observable, circumpolar,
gamma-ray--quiet sources. Of these, 17 sources were observed (71\,\%), in order of
decreasing polar distance, until the end of our June survey. Since the polar distance criterion
is independent of source properties, the resulting gamma-ray--quiet sources is again an
unbiased subsample of the statistically complete sample of 25 sources. This unbiased
subsample is 86\,\% ($6/7$) complete for sources with $R$-mag $\leq16$, 92\,\% ($11/12$)
for sources with $R$-mag $\leq17$ and of course 71\,\% ($17/24$) for sources with
$R$-mag $\leq18$.

Our gamma-ray--loud sources are (by construction of CGRaBS) similar in
radio and X-ray flux, and radio spectra. Their
R-Magnitudes span a similar range and have a similar distribution as
can be seen a posteriori (see Fig.~\ref{photometry}). The radio
modulation indices have been shown to be systematically higher for
gamma-ray--loud sources in general \cite{RichardsEtal2011}; to counter this
effect, we select, among the gamma-ray--quiet candidates, only sources with
statistically significant radio variability, as quantified by the
modulation index (see table~\ref{sample_select}). The gamma-ray--loud
sample has fractionally more BL Lacs (about 50\%) than the
gamma-ray--quiet control sample (about 10\%), which is expected when 
comparing gamma-ray--loud with gamma-ray--quiet samples, but which
should, however, be taken into account when interpreting our results. 

The June 2013 survey results are discussed in Section~\ref{observations}. 

\subsection{Monitoring Sample}

The data collected during the survey phase were used for the construction of the 2013
observing season monitoring sample which was observed from July 2013 until the end of the
2013 observing season, with an approximate average cadence of once every 3 days. It
consists of three distinct groups.
\begin{enumerate}
\item An unbiased subsample of a statistically complete sample of gamma-ray--loud blazars.
  Starting from the ``parent sample'' and applying the selection criteria summarised in
  Table~\ref{sample_select}, we obtain a statistically complete sample of 59
  sources. Application of field-quality cuts (based on data from the June survey) and
  location-on-the-sky criteria that optimise continuous observability results in an
  unbiased subsample of 51
  sources. 
\item An unbiased subsample of a statistically complete sample of gamma-ray--quiet
  blazars. Starting from the CGRaBS, excluding sources in the 2FGL, and applying the
  selection criteria summarised in Table~\ref{sample_select} results in a statistically
  complete sample of 22 sources. Our ``control sample'' is then an unbiased subsample of
  10 blazars, selected from this complete sample of gamma-ray--quiet blazars with
  field-quality and location-on-the-sky criteria.
\item 24 additional ``high interest'' sources, that did not otherwise make it to the
  sample list.
\end{enumerate}

These observations will later allow the characterisation of each source's typical
behaviour (i.e. average optical flux and degree of polarization, rate of change of
polarization angle, flux and polarization degree variability characteristics). This
information will be further used to:
\begin{enumerate}
\item improve the optical polarization parameters estimates for future polarization
  population studies,
\item develop a dynamical scheduling algorithm, aiming at self-triggering higher cadence
  observing for blazars displaying interesting polarization angle rotation events, for the
2014 observing season, 
\item improve the definition of our 2014 monitoring sample using a contemporary average, rather
  than archival single-epoch, optical flux criterion along with some estimate of the
  source variability characteristics in total intensity and in polarized emission.
\end{enumerate}

For the June survey control sample sources, circumpolar sources were selected so that gamma-ray--quiet source observations could be taken at any time and the gamma-ray--loud sources could be prioritized. In contrast, the gamma-ray--quiet sources for the monitoring sample were selected in order of {\em increasing} declination, to avoid as much as possible the northernmost sources which suffer from interference in observations by strong northern winds at times throughout the observing season at Skinakas.

\begin{figure*}
\includegraphics[width=0.9\textwidth,clip=true]{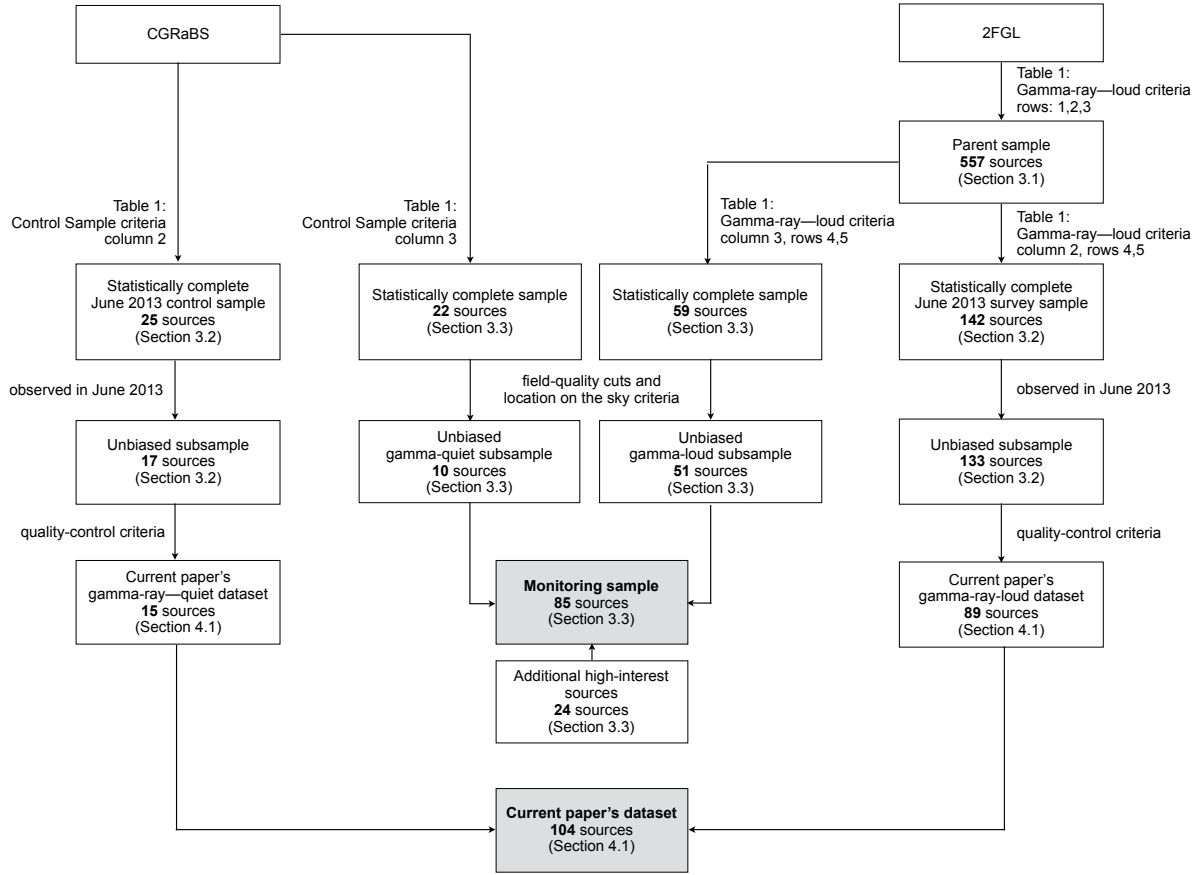}
\caption{\label{chart} Flow chart indicating the steps for the selection of (a)
 {\em left-hand half}: the gamma-ray--quiet ``control'' sample and 
 (b) {\em right-hand half}: the gamma-ray--loud sample, separately for the monitoring
 program of 2013 and the current paper's study.}
\end{figure*}

The steps followed for the selection of the June survey sample and the first season
monitoring one are summarised schematically in Fig.~\ref{chart}. The complete sample of
our monitored sources is available at {\tt robopol.org}.

The 2nd Fermi gamma-ray catalog, from which the parent gamma-ray-loud sample is drawn,
represents a ``from-scratch'' all-sky survey in $\ge100$~MeV gamma-rays, so the limit in
gamma-ray flux should in principle result in a clean, flux-limited sample. One possible
source of bias however is the process of characterisation of a source as a ``blazar'', in
which case other catalogs of blazars (principally from radio surveys) are used for the
identification and classification of sources. Given that 575 of the 1873 Fermi catalog
sources are unassociated, that bias may in fact be non-trivial: we don't know how many of
the unassociated sources are blazars, and we cannot a priori be certain that the
properties of any blazars among the unassociated sources are similar to those of
confidently associated blazars. Unassociated sources are not however uniformly distributed
among fluxes and Galactic latitudes. Brighter sources, sources in high Galactic latitudes,
and sources with hard spectra tend to have smaller positional error circles and are more
easily associated with low-energy counterparts (because of more photons available for
localization, lower background, and better single-photon localization at higher energies,
respectively). The first of these two factors lower the fraction of unassociated sources
among the 2FGL sources that satisfy our Galactic latitude and gamma-ray--flux cuts, from
$\sim30\%$ to $\sim 20\%.$ The effect of possible biases due to the presence of
unassociated sources in 2FGL can be further assessed, as part of theoretical population
studies, under any particular assumption regarding the nature of these sources, as well as
if, at some point in the future, a large fraction of these sources become confidently
associated with low-energy counterparts.

Any other minor biases entering through our choices of limits in
gamma-ray flux and $R$ magnitude can be accounted for in theoretical
population studies given the cuts themselves, the uncertainty
distribution in the measured quantities (which can be found in the
literature), and some knowledge of the
variability of these sources (obtainable from Fermi data in gamma
rays, and, at the most basic level, from comparing
historical magnitudes with magnitudes from this work in the R-band). 

\section{Results of June 2013 Survey}\label{observations}

\subsection{Observations}
The June survey observations took place between June 1st and June 26th. During that
period, we conducted RoboPol observations, weather permitting, for 21 nights. Of those, 14
nights had usable dark hours. The most prohibiting factors have been wind, humidity and
dust, restricting the weather efficiency to 67\,\%, unusually low based on historical
Skinakas weather data. During this period, a substantial amount of observing time was
spent on system commissioning activities. In the regular monitoring mode of operations a
much higher efficiency is expected.

During the survey phase a total of 135 gamma-ray--loud targets (133 of them comprising the
unbiased subsample of the 142-source sample and 2 test targets), 17 potential
control-sample sources, and 10 polarization standards, used for calibration purposes, were
observed. For the majority of the sources, a default exposure time of $15-17.5$ min
divided into 3 exposures was used to achieve a polarization sensitivity of ${\rm SNR}_{p}
= 10:1$ for a 17 mag source with polarization fraction of 0.03, based on the instrument
sensitivity model. Shorter total exposures were used for very bright sources and standard
stars, and their duration was estimated on-the-fly. Typically we observed 2 different
polarimetric standards every night to confirm the stability of the instrument (see
``pipeline'' paper).  

In summary, we observed 133 + 17 blazars belonging to the unbiased subsamples of the gamma-ray--loud and gamma-ray--quiet complete samples respectively. Of these sources, 89 gamma-ray--loud and 15
gamma-ray--quiet sources passed a series of unbiased, source-property--independent quality-control criteria to ensure accurate
polarization measurements (see Fig.~\ref{chart}).

 The RoboPol results for these 89 + 15 sources are shown in
 Table~\ref{SurveyResults}. These results include: the $R$-magnitude, calibrated with two
 different standards [the Palomar Transient Factory (PTF)  $R$-band catalog \citep[][whenever
 available]{2012PASP..124..854O} or the USNO-B catalog \citep{Monet:2003p2377}]; the
 polarization fraction, $p = \sqrt{u^2 + q^2}$; and the polarization angle,
 $\chi=\frac{1}{2}\arctan\left(\frac{u}{q}\right),$ measured from the celestial north
 counter-clockwise. In that table target sources are identified by the prefix ``RBPL'' in
 their RoboPol identifying name. Additional archival information for
 these sources are given in Table 3, available as online as supplementary
 material. 

The images were processed using the data reduction pipeline described in the ``pipeline''
paper. The pipeline performs aperture photometry, calibrates the measured counts according
to an empirical instrument model, calculates the linear polarization fraction $p$ and angle
$\chi$, and performs relative photometry using reference sources in the frame to obtain
the $R$-band magnitude. Entries in table~\ref{SurveyResults} with {\em no} photometry
information are sources for which PTF data do not exist and the USNO-B data were not of
sufficient quality for relative photometry. Polarimetry, for which only the relative
photon counts in the four spots are necessary, can still of course be performed without
any problem in these cases. The photometry error bars are dominated by uncertainties in
our field standards, while the polarization fraction and angle errors are photon-count
dominated. For the few cases where multiple observations of a source were obtained in
June, weighted averaging of the $q$ and $u$ has been performed. The quoted uncertainty
follows from formal error propagation assuming that $q$ and $u$ follow normal distributions 
and that the polarization has not changed significantly between measurements.

\subsection{Debiasing}\label{debiasing} 
The $p$ values and uncertainties $\sigma_p$ shown in Table~\ref{SurveyResults} are the raw
values as produced by the pipeline, without any debiasing applied to them, and without
computing upper limits at specific confidence levels for low $p/\sigma_p$
ratios. Debiasing is appropriate for low signal-to-noise measurements of $p$ because
measurements of linear polarization are always positive and for any true polarization degree
$p_0$ we will, on average, measure $p > p_0$.  \citet{john} gives approximations for the
maximum-likelihood estimator of $p_0$ at various $p/\sigma_p$ levels, and describes how to
calculate appropriate upper limits for specific confidence levels. He finds that the
maximum-likelihood estimator is well approximated by 
\begin{equation}
\hat{p} = \left\{ \begin{array}{lll}
 0 & &\mbox{for $p/\sigma_p<\sqrt{2}$} \\
  \sqrt{p^2-\sigma_p^2} & &\mbox{for $p/\sigma_p\gtrsim3$}
       \end{array} \right.
\end{equation}
For $p/\sigma_p\gtrsim 3$ the assumption of a normal distribution for $p-$measurements is
also acceptable (and it is a good assumption for $p/\sigma_p \gtrsim 4$). Debiasing is not
necessary for polarization angles $\chi$, as the most probable measured value is the true
$\chi$ and as a result the pipeline output is an unbiased $\chi$ estimator.

Whenever in the text debiased $p$ values are mentioned, we are referring to a correction
using $p_{\rm debiased}\approx \sqrt{p^2-\sigma_p^2}$ down to $p/\sigma_p =\sqrt{2}$ and 0
for lower signal-to-noise ratios (a choice frequently used in the literature), despite the
fact that below $p/\sigma_p\sim 3$ this recipe deviates from the maximum-likelihood
estimator.  When a good estimate of the uncertainty is also necessary (i.e. in our
likelihood analyses), we only use measurements with $p/\sigma_p >3$, for which not only
the debiasing recipe we use is close to the maximum-likelihood estimator, but also
the uncertainty calculated by the pipeline $\sigma_p$ is a reasonable approximation to the
$68\%$ uncertainty in the value of $p$.

\subsection{Polarization properties of gamma-ray--loud vs gamma-ray--quiet blazars}\label{KSTest}
\begin{figure}
\includegraphics[width=8cm,clip=true]{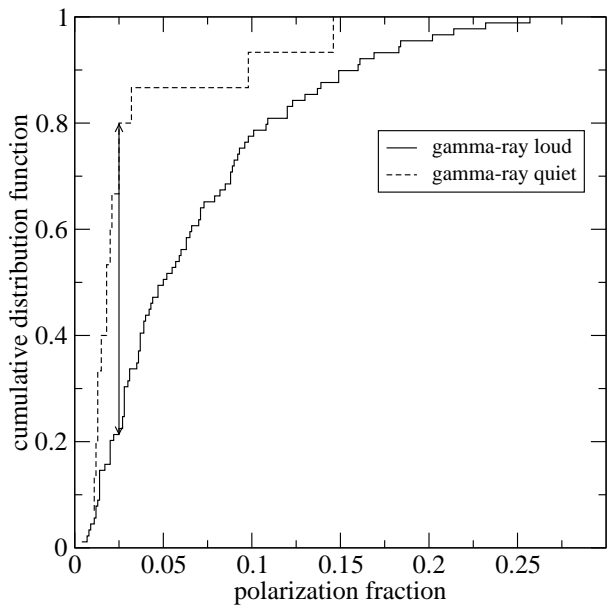}
\caption{\label{fig_KS_main} Cumulative distribution functions of raw $p$ values for all 89 gamma-ray--loud blazars (solid line) and 15 gamma-ray--quiet blazars (dashed line) with observations in June 2013 that passed all our quality cuts. The maximum difference between the two ($=0.6$) is shown with the double arrow. The hypothesis that the two samples are drawn from the same distribution is rejected at the $4\times10^{-4}$ level ($3.5\sigma$).}
\end{figure}

As the unbiased nature of our samples allows us to address issues related to the blazar population, we wish to ask the question: are the measured polarization fractions of gamma-ray--loud and gamma-ray--quiet blazars consistent with having been drawn from the same distribution?

Because our observing strategy and data processing pipeline is uniform across sources, if the intrinsic polarization fractions of gamma-ray--loud and gamma-ray--quiet sources were indeed drawn from the same distribution, then the resulting {\em observed} distributions of $p$  would also be consistent with being the same. Each of them might not be consistent with the intrinsic $p$ distribution of the blazar population, because of biasing, and because at low $p/\sigma_p$ values what is being recorded is in general more noise than information; however, biasing and noise would affect data points in both populations in the same way and at the same frequency, and the resulting observed distributions, no matter how distorted, would be the same for the two subpopulations. 

For this reason, we compare the observed raw $p-$values (as they come
out of the pipeline) of the two samples of 89 gamma-ray--loud sources
and 15 gamma-ray--quiet sources.  Figure~\ref{fig_KS_main} shows the
cumulative distribution  functions (CDFs) of raw $p$ values for the
gamma-ray--loud blazars (solid line) and the gamma-ray--quiet blazars
(dashed line).  The maximum difference between the two CDFs (indicated
with the double arrow) is $0.58$, and a two-sample Kolmogorov-Smirnov
test rejects the hypothesis that the two samples are drawn from the
same distribution at the $4\times10^{-4}$ level ($3.5\sigma$). The observed
raw $p-$distributions are therefore inconsistent with being identical,
and, as a result, the underlying distributions of intrinsic $p$ cannot
be identical either.

As discussed in \S  \ref{jss}, while the gamma-ray--quiet sample is a
pure subsample of CGRaBS, the gamma-ray--loud sample contains many
(47) non-CGRaBS sources, which, in practice, means that the fraction
of BL Lac objects (bzb) is much higher. To test whether this is the
source of the discrepancy, we have repeated the same test between the
42 CGRaBS sources in our gamma-ray--loud sample, and the 15 sources in
our gamma-ray--quiet sample. The maximum difference between the two
CDFs in this case is $0.54$, so the hypothesis that the two
distributions are identical is again rejected at the $3\times10^{-3}$ level
($3\sigma$).

We conclude that {\em the optical polarization properties of gamma-ray--loud and gamma-ray--quiet sources are different}. 

\subsection{Polarization fraction vs $R$ magnitude}\label{pvsR}

\begin{figure}
\includegraphics[width=8cm,clip=true]{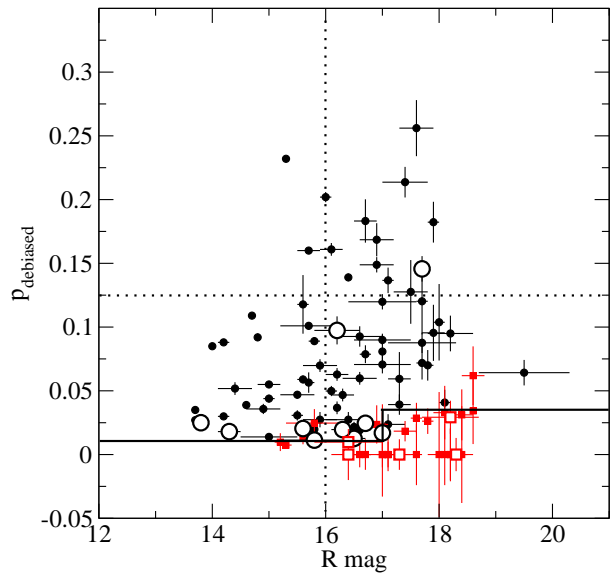}
\caption{\label{photometry} Debiased polarization
  fraction versus $R$ magnitude. Above the solid line, we assume that we can measure
  polarization independently of source brightness (for details see
  Sect.~\ref{observations}). Black circles correspond to measurements with $p/\sigma_p\geq
  3$, and red squares to measurements with $p/\sigma_p< 3$. Filled symbols correspond to
  gamma-ray--loud sources and open symbols to gamma-ray--quiet sources.  }
\end{figure}

We next turn our attention to the behaviour of the polarization fraction with $R$ magnitude. In Fig.~\ref{photometry}
we plot the debiased value of the polarization fraction as a function of the measured $R$ magnitude for each source. 
Sources for which $p/\sigma_p<3$ are shown with red colour. There are two noteworthy pictures in this plot: the clustering of low signal-to-noise ratio measurements in the lower-right corner of the plot, and the scarcity of observations in the upper-left part of the plot. 

The first effect is expected, as low polarization fractions are harder to measure for
fainter sources with fixed time integration. This is a characteristic of the June survey
rather than the RoboPol program in general: in monitoring mode, RoboPol scheduling
features adaptive integration time to achieve a uniform signal-to-noise ratio down to a
fixed polarization value for any source brightness. For source brightness higher than
magnitude of 17 we have measured polarization fractions down to $1.5\times 10^{-2}$: most
measurements at that level have $p/\sigma_p\geq 3$. For source brightness lower than
magnitude of 17 the same is true for polarization fractions down to $3.5\times
10^{-2}$. These limits are shown with the thick solid line in Fig.~\ref{photometry}, and
they are further discussed in Section~\ref{shapes} in the context of our likelihood
analysis to determine the most likely intrinsic distributions of polarization fractions
for gamma-ray--loud and gamma-ray--quiet sources.

The second effect -- the lack of data points for $R$ magnitudes lower than 16 and polarization fractions higher than $1.25 \times 10^{-1}$ as indicated by the dotted lines -- may be astrophysical in origin: in sources where unpolarized light from the host galaxy is a significant contribution to the overall flux, the polarization fraction should be on average lower. This contribution also tends to make these sources on average brighter. We will return to a quantitative evaluation and analysis of this effect when we present data from our first season of monitoring, using both data from the literature as well as our own variability information to constrain the possible contribution from the host for as many of our sources as possible.

\subsection{Intrinsic distributions of polarization fraction}\label{shapes}
\begin{figure}
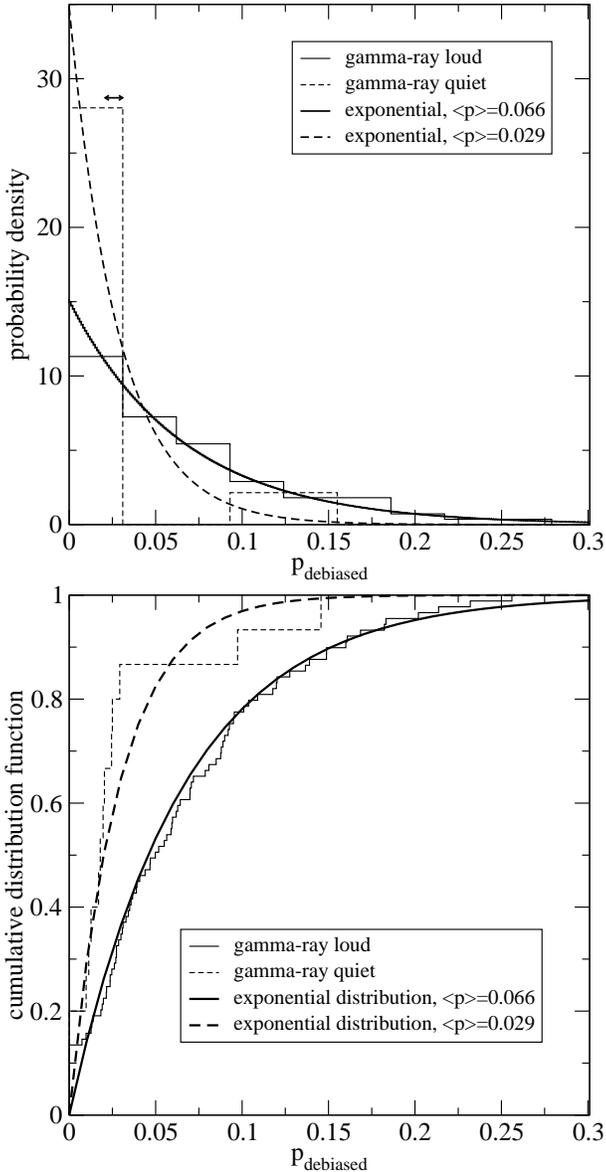

\includegraphics[width=8cm,clip=true]{polarimetry_review.eps}
\includegraphics[width=8cm, clip]{cdf_debiased_review.eps}
\caption{\label{fig1} Histogram (upper panel) and cumulative distribution function (lower panel) of debiased $p$ values for all 89 gamma-ray--loud (thin solid lines) and 15 gamma-ray--quiet (dashed lines) blazars that pass quality cuts. The typical measurement uncertainty is shown in the upper panel with the arrow; the uncertainty spread is $\sim 10\%$ of that value. Thick solid and dashed lines correspond to the PDF and CDF of exponential distributions with average equal to the sample average of each population. }
\end{figure}

In Section \ref{KSTest} we showed that the intrinsic distributions of polarization fraction of gamma-ray--quiet and gamma-ray--loud blazars must be different; however, that analysis did not specify what these individual intrinsic distributions might be. We address this issue in this section. Our approach consists of two steps. First, we will determine what the overall shape of the distributions looks like, and we will thus select a {\em family of probability distribution functions} that can best describe the intrinsic probability distribution of polarization fraction in blazars. Next, we will use a likelihood analysis to produce best estimates and confidence limits on the parameters of these distributions for each subpopulation. 

\subsubsection{Selection of Family of Distributions}

In order to determine the family of distributions most appropriate to describe the
polarization fraction of the blazar population we plot, in the upper panel of
Fig.~\ref{fig1}, a histogram -- normalised so that it represents a probability density --
of all the debiased $p$ values in the gamma-ray--loud and gamma-ray--quiet samples, independently of
their $p/\sigma_p$ ratio (89 and 15 sources, respectively). It appears that these histograms resemble exponential distributions. 
Indeed, in the upper panel of Fig.~\ref{fig1}, we also over-plot the
exponential distributions with mean equal to the sample average of $p$ for each
sample, and we see that there is good agreement in both cases. To verify that our choice of binning does not affect the appearance of these distributions, we also plot, in the lower panel of Fig.~\ref{fig1}, the CDF of each sample, as well as the CDFs corresponding to each of the model PDFs in the upper panel. The agreement is again excellent. We conclude that {\em the PDFs of the polarization fraction of gamma-ray--quiet and gamma-ray--loud blazar subpopulations can be well described by exponential distributions.}

\subsubsection{Determination of Distribution Parameters}
\begin{figure}
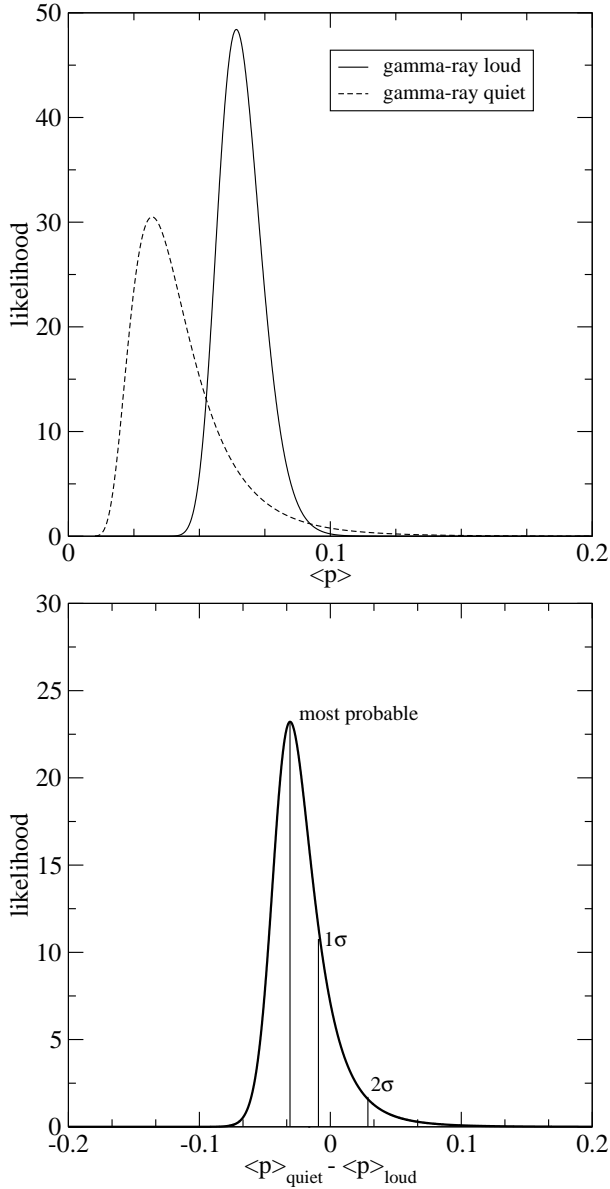

\includegraphics[width=8cm, clip]{likes_review.eps}
\includegraphics[width=8cm,clip=true]{CC_review.eps}
\caption{\label{fig2} Likelihood of $\left<p\right>$ for each population (upper panel) and of the
  difference of $\left<p\right>$ between the two populations (lower panel). The most-probable values
  differ by about a factor of 2. }
\end{figure}

In this section, we seek to determine the best estimate values and associated confidence intervals for the  parameters of the intrinsic PDFs of polarization fraction for our two blazar subpopulations. All values of $p$ used in this section are debiased as described in Section \ref{debiasing}.
Based on the results of our previous discussion, we will  assume that the probability distribution of $p$ in a sample of blazars can be described as
\begin{equation}\label{thedist}
  P(p) dp = \frac{1}{\pmean } \exp \left( -\frac{p}{\pmean}\right) dp\,. 
\end{equation}
In order to be formally correct, there should be a factor of $1-{\rm e}^ {-1/\pmean}$ in
the denominator of Equation \ref{thedist} to correct for the fact that $p$ is defined in
the $[0,1]$ rather than the $[0,\infty)$ interval; the correction is however small
for the values of $\pmean$ that are of interest here.  The mean, $\pmean$, is the single
parameter of this family of distributions, and it is the quantity that we seek to estimate
from our data for each subsample.

In the population studies that follow, we will include only sources with $p/\sigma_p\geq
3$. However, in order to avoid biasing our statistics by this choice, we apply
sharp cuts in $p-$space that exclude most, if not all, of our low $p/\sigma_p$
measurements; these cuts can then be explicitly corrected for in our analysis (which will
assume that sources below a certain $p$ value do exist, in numbers predicted by the
exponential distribution, but cannot be measured). These selection criteria are visualised
by the thick solid  line in Fig.~\ref{photometry}. 

We thus split each population into two sub-samples, along the (measured) $R$-mag 17 line,
and, for each population, we consider each subsample to be a distinct ``experiment'' with
a different data cut ($1.5\times 10^{-2}$ for bright sources and $3.5\times 10^{-2}$ for
faint sources). We then use a likelihood analysis to estimate the maximum-likelihood value
of the average $\pmean$ for each population, in a fashion similar to the one implemented
for population studies in \cite{RichardsEtal2011}. The sources for which no photometry
information is available are considered part of our second experiment and the stricter cut
is applied to them. In all our calculations below we use debiased values of $p$.

The likelihood of a single observation of a polarization fraction $p_i$ of (approximately)
Gaussian uncertainty $\sigma_i$ drawn from the distribution of Eq.~(\ref{thedist}) with
mean $\pmean$ can be approximated by 
\begin{eqnarray}
  \label{eq:single_likelihood}
  \ell_i &=& \!\! \!\!\!\! \int_{p=0}^{\infty} dp \frac{1}{\pmean} \exp\left(-\frac{p}{\pmean}\right)
  \frac{\exp\left[-\frac{(p-p_i)^2}{2\sigma_i^2}\right]}{\sigma_i\sqrt{2\pi}} 
  \nonumber \\
  &=& \frac{1}{2\pmean}
  \exp\left[-\left(\frac{p_i}{\pmean}-\frac{\sigma_i^2}{2\pmean^2 }\right)\right]\times \nonumber \\
  && \left[1+{\rm erf}\left(\frac{p_i}{\sigma_i\sqrt{2}}-\frac{\sigma_i}{\sqrt{2}\pmean}\right)\right].\nonumber \\
\end{eqnarray}
Extending the upper limit of integration to $\infty$ instead of 1 simplifies the mathematics while introducing no appreciable change in our results, as the exponential distribution approaches 0 fast at $p<1$ for the data at hand. This can be directly seen in Fig.~\ref{fig1}. 

In order to implement data cuts restricting $p_i$ to be smaller than some limiting value
$p_l$, the likelihood of a single observation $p_i$ will be given by Eq.~\ref{eq:single_likelihood}
multiplied by a Heaviside step function, and re-normalised so that the likelihood
$\ell_{i,\rm cuts}$ to obtain any value of $p_i$ {\em above} $p_l$ is 1:
\begin{equation}
  \ell_{i,\rm cuts}\left(p_l\right) = \frac{H(p_i-p_l)\ell_i}{\int _{p_i=p_l}^1 dp_i \ell_i}\,.
\end{equation}
This re-normalisation ``informs'' the likelihood that the reason why no observations of
$p_i<p_l$ are made is not because such objects are not found in nature, but rather because
we have excluded them ``by hand.'' We are, in other words, only sampling the $p>p_l$ tail
of an exponential distribution of mean $\pmean$.
The likelihood of $N$ observations of this type is
\begin{equation}\label{likelihood0}
  \mathcal{L} (\pmean) = \prod_{i=1}^N \ell_{i,\rm cuts}\left(p_l\right)\,,
\end{equation}
and the combination of two experiments with distinct data cuts, described above, will have a likelihood equal to 
\begin{equation}\label{likelihood}
  \mathcal{L} (\pmean)= \prod_{i=1}^{N_l} \ell_{i,\rm cuts}\left(p_l\right)\prod_{j=1}^{N_u} \ell_{j,\rm cuts}\left(p_u\right)\,,
\end{equation}
where $N_l$ (equal to 42 for the gamma-ray--loud sources and 6 for the gamma-ray--quiet
sources) is the number of $p/\sigma_p>3$ objects with $R$-mag $<17$ surviving the $p_l=0.015$ cut, and $N_u$
(equal to 21 for the gamma-ray--loud sources and 1 for the gamma-ray--quiet sources) is
the number of  $p/\sigma_p>3$ objects with $R$-mag $>17$ or no photometry information surviving the $p_u=0.035$ cut.  Maximising
Eq. (\ref{likelihood}) we obtain the maximum-likelihood value of $\pmean$.  Statistical
uncertainties on this value can also be obtained in a straight-forward way, as
Eq. (\ref{likelihood}), assuming a flat prior on $\pmean$, gives the probability density
of the mean polarization fraction $\pmean$ of the population under study.

The upper panel of Fig.~\ref{fig2} shows the likelihood of $\pmean$ for the
gamma-ray--loud (solid line) and gamma-ray--quiet (dashed line) populations. The
maximum-likelihood estimate of $\pmean$ with its 68\% confidence intervals is $6.4
^{+0.9}_{-0.8}\times10^{-2}$ for gamma-ray--loud blazars and $3.2^{+2.0}_{-1.1}\times10^{-2}$ for
gamma-ray--quiet blazars. The maximum-likelihood values of $\pmean$ differ by
more than a factor of 2, consistent with our earlier finding that the two populations have different polarization fraction PDFs. However, because of the small number of gamma-ray--quiet sources surviving the strict signal-to-noise cuts we have imposed in this section (only 7 objects), the gamma-ray--quiet $\pmean$ cannot be pinpointed with enough accuracy and its corresponding likelihood exhibits a long tail towards high values. 
For this reason, the probability distribution of the difference between the $\pmean$ of the two populations, which is quantified by  the cross-correlation of the two likelihoods, has a peak, at a difference of $3.1\times 10^{-2}$, which is less than  2$\sigma$ from zero. This result is shown in the lower panel of Fig.~\ref{fig2}.

The accuracy with which the gamma-ray--quiet $\pmean$ is estimated can
be improved in two ways. First, by an improved likelihood analysis
which allows us to properly treat even low $p/\sigma_p$ sources (e.g.,
\citealp{ss85,john}). And second, by an improved survey of the
gamma-ray--quiet population (more sources to improve sample
statistics, and longer exposures to improve the accuracy of individual
$p$ measurements). A more difficult-to-assess uncertainty, especially
at low values of $p$, is the effect of interstellar
polarization. However, because of the ability of the RoboPol
instrument to measure polarization properties for all sources in its
large $13'\times 13'$ field of view, the amount of
interstellar-dust--induced polarization can in principle be estimated
studying the polarization properties of field stars in the vicinity of
each blazar. We will return to this problem in the future, with 
further analysis of our already-collected data.

\subsection{Polarization angles}
\begin{figure}
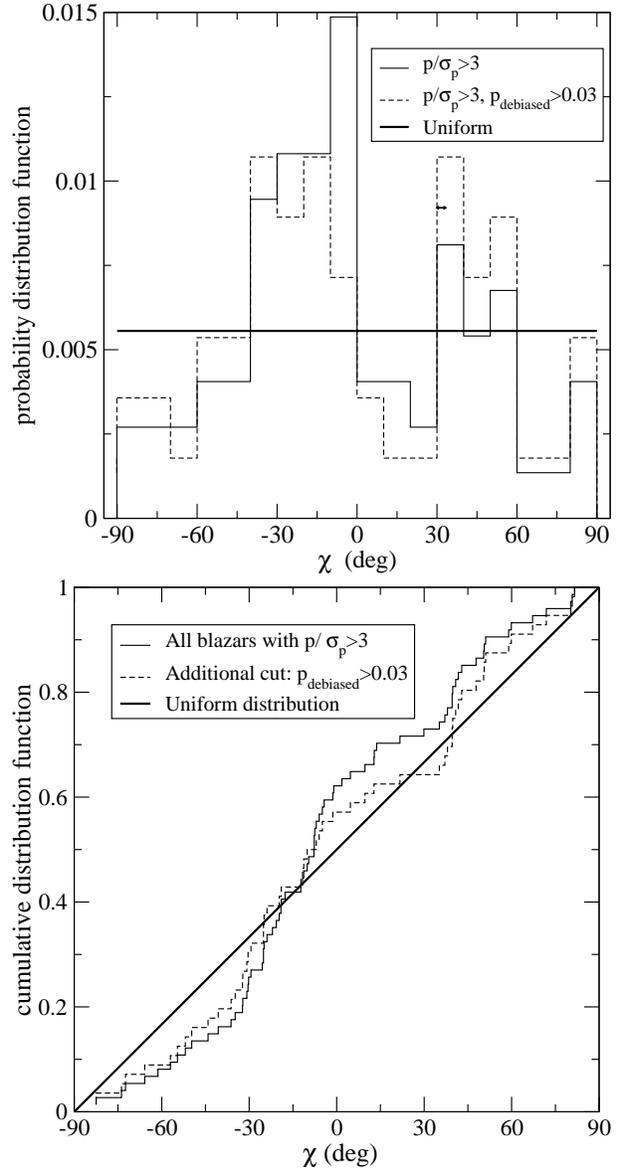

\includegraphics[width=8cm, clip]{pdf_review.eps}
\includegraphics[width=8cm,clip=true]{cumulative_review.eps}
\caption{\label{fig3}Upper panel: histogram of polarization angles $\chi$ for all sources with $p/\sigma_p\geq 3$. The double arrow 
represents the typical uncertainty on $\chi$; the associated spread in uncertainties is about $10\%$ of that value. Lower panel: cumulative 
distribution of the polarization angles $\chi$. The PDF and CDF of a uniform $\chi$ distribution are over-plotted in the 
upper and lower panel respectively with the thick solid line.}
\end{figure}

In this section we assess the consistency of the measured polarization
angles, $\chi$, with an expected uniform distribution. For this
reason, we plot in  Fig.~\ref{fig3}, the histogram (normalised so that it corresponds to a PDF) and the CDF of the polarization angles $\chi$, for all 
sources with $p/\sigma_p \ge 3$ (72 sources). 
The difference from the (overplotted) uniform distribution is not statistically significant: the maximum difference between the two CDFs is 0.132 and a Kolmogorov--Smirnov test finds the two distributions consistent at the 15\,\% level. The agreement further improves to the $1\sigma$ level if we only include sources that satisfy the additional requirement that $p>3\times10^{-2}$ (56 sources, dashed lines in Fig.~\ref{fig3}). 

The reason for the difference between the uniform distribution and
that of the measured $\chi$ when sources with low (but
high-significance) $p-$values are included is likely
astrophysical. For low polarization sources, any foreground
polarization picked up by their optical light during propagation would
be a larger fraction of the overall polarization, and any preferred
direction in the foreground polarization would affect more
significantly the final value of $\chi$. Indeed, half of the sources
removed by the $p=3\times 10^{-2}$ have polarization angles covering
only a small range of values, between $-20$ and $0$ degrees (close to the maximum of the solid-line histogram in the upper panel 
of Fig.~\ref{fig3}.) This is exactly the behaviour that would be
expected from low-level foreground polarization in a preferred
direction (see discussion in \S~\ref{discussion}).

\begin{table*}
 \centering
\scriptsize
\caption{\label{SurveyResults} Photometric and polarization results of the RoboPol June
  2013 optical polarization survey of gamma-ray--loud blazars and the gamma-ray--quiet
  control sample sources.  Note that the polarization angle $\chi$ has been corrected for
instrumental rotation of 2.31$\pm$0.34\,deg as discussed in ``pipeline'' paper. The error
 in the instrumental rotation is accounted for in the final error
 estimate through formal error  propagation. This table is also
available electronically as supplementary material.}
  \begin{tabular}{l r@{\,$\pm$\,}l r@{\,$\pm$\,}l r@{\,$\pm$\,}l r l r@{\,$\pm$\,}l r@{\,$\pm$\,}l r@{\,$\pm$\,}l r}
  \hline
RoboPol ID &\multicolumn{2}{c}{$R^1$} &\multicolumn{2}{c}{$p$}        &\multicolumn{2}{c}{$\chi$} &\multicolumn{1}{c}{Date$^4$} & RoboPol ID &\multicolumn{2}{c}{$R^1$} &\multicolumn{2}{c}{$p$}     &\multicolumn{2}{c}{$\chi$} &\multicolumn{1}{c}{Date$^4$} \\
           &\multicolumn{2}{c}{(mag)} &\multicolumn{2}{c}{(fraction)} &\multicolumn{2}{c}{(deg)}  &\multicolumn{1}{c}{} &             &\multicolumn{2}{c}{(mag)} &\multicolumn{2}{c}{(fraction)} &\multicolumn{2}{c}{(deg)} &\multicolumn{1}{c}{} \\
 \hline
\\\multicolumn{16}{c}{\bf Target Sample}\\\\[-1.5ex]
RBPL\,J0841$+$7053    &16.6  &0.1$^{2}$  &0.020 &0.006  &$-$19.1 &9.0&J24		&RBPL\,J1637$+$4717 &18.1  &0.1  &0.042 &0.010  &$-$12.3 &6.5  &J01, J06     \\ 
RBPL\,J0848$+$6606 &18.2  &0.1$^2$ &0.014 &0.021  &8.2  &43.3  &J21		&RBPL\,J1642$+$3948 &17.6  &0.1  &0.031 &0.012  &61.9 &10.9 &J08		       \\ 
RBPL\,J0956$+$2515 &17.6  &0.1     &0.020 &0.024  &$-$65.3 &36.9  &J10		&RBPL\,J1643$-$0646 &16.9  &0.2  &0.028 &0.015  &14.1 &15.1 &J08		       \\ 
RBPL\,J0957$+$5522 &15.7  &0.1     &0.057 &0.008  &4.7  &4.2&J25		&RBPL\,J1649$+$5235 &17.0  &0.5  &0.090 &0.004  &$-$30.8 &1.3  &J09		       \\ 
RBPL\,J1014$+$2301 &17.0  &0.1     &0.009 &0.010  &30.1 &31.3  &J11		&RBPL\,J1653$+$3945 &13.7  &0.02$^2$&0.027 &0.001  &1.8  &0.8  &J01, J27     \\ 
RBPL\,J1018$+$3542 &17.1  &0.1     &0.014 &0.013  &$-$61.5 &28.1  &J19		&RBPL\,J1722$+$1013 &17.6  &0.3  &0.257 &0.022  &$-$30.4 &2.8  &J10, J27     \\
RBPL\,J1032$+$3738 &17.9  &0.2     &0.098 &0.022  &47.8 &6.3&J25		&RBPL\,J1727$+$4530 &17.3  &0.2  &0.063 &0.021  &50.4 &9.1  &J10		       \\ 
RBPL\,J1037$+$5711 &16.2  &0.04    &0.037 &0.005  &42.9 &3.6&J22, J24	&RBPL\,J1748$+$7005 &15.7  &0.2  &0.160 &0.002  &67.2 &0.5  &J19		       \\ 
RBPL\,J1041$+$0610 &16.7  &0.1$^2$ &0.011 &0.012  &$-$54.4 &21.2  &J08		&RBPL\,J1749$+$4321 &17.4  &0.4  &0.214 &0.012  &$-$1.3  &1.6  &J09		       \\ 
RBPL\,J1048$+$7143 &15.9  &0.3$^2$ &0.070 &0.005  &$-$44.1 &2.1&J24		&RBPL\,J1754$+$3212 &16.6  &0.3  &0.060 &0.005  &$-$6.0  &2.5  &J18		       \\ 
RBPL\,J1054$+$2210 &17.7  &0.1$^3$ &0.073 &0.013  &$-$4.9  &5.4&J08		&RBPL\,J1800$+$7828 &16.3  &0.2  &0.047 &0.005  &$-$72.4 &3.2  &J21		       \\ 
RBPL\,J1058$+$5628 &14.9  &0.3$^3$ &0.036 &0.004  &$-$57.0 &2.9&J21&RBPL\,J1806$+$6949 &14.2  &0.1  &0.088 &0.002  &81.6 &0.6  &M26, J19     \\ 
RBPL\,J1104$+$0730 &\mc{2}{c}{\ldots} &0.149 &0.007  &38.0 &1.4&J09, J25	&RBPL\,J1809$+$2041 &19.5  &0.8  &0.065 &0.010  &$-$32.2 &4.5  &J08		       \\ 
RBPL\,J1121$-$0553 &18.4  &0.1$^2$ &0.037 &0.038  &$-$49.4 &30.7  &J09		&RBPL\,J1813$+$0615 &16.1  &0.2  &0.161 &0.005  &39.6 &1.0  &J12, J24     \\ 
RBPL\,J1132$+$0034 &17.8  &0.1     &0.071 &0.012  &$-$82.5 &4.7&J11, J26	&RBPL\,J1813$+$3144 &16.1  &0.1  &0.050 &0.004  &51.1 &2.2  &J03, J24     \\ 
RBPL\,J1152$-$0841 &18.0  &0.2     &0.007 &0.049  &$-$62.6 &197.1 &J11		&RBPL\,J1824$+$5651 &15.5  &0.1  &0.031 &0.004  &$-$51.9 &3.8  &M31, J09     \\ 
RBPL\,J1203$+$6031 &15.6  &0.04$^2$&0.014 &0.005  &38.8 &10.6  &J19		&RBPL\,J1836$+$3136 &17.0  &0.6  &0.120 &0.006  &$-$23.9 &1.4  &J08		       \\ 
RBPL\,J1204$-$0710 &16.4  &0.4     &0.028 &0.006  &$-$7.8  &9.2&J06		&RBPL\,J1838$+$4802 &15.6  &0.1$^2$ &0.059 &0.003  &37.1 &1.3  &J03		       \\ 
RBPL\,J1217$+$3007 &14.7  &0.02$^2$&0.109 &0.002  &$-$11.7 &0.5&J22, J24	&RBPL\,J1844$+$5709 &17.3  &0.2  &0.040 &0.008  &$-$49.7 &5.9  &J22		       \\ 
RBPL\,J1220$+$0203 &15.3  &0.1     &0.008 &0.003  &11.0 &10.0  &J23		&RBPL\,J1849$+$6705 &18.6  &0.2  &0.066 &0.023  &13.5 &10.1 &J20		       \\ 
RBPL\,J1222$+$0413 &18.0  &0.1$^2$ &0.108 &0.030  &$-$91.2 &5.9&J23		&RBPL\,J1903$+$5540 &15.7  &0.5  &0.101 &0.003  &41.7 &0.9  &J22		       \\ 
RBPL\,J1224$+$2436 &15.8  &0.1$^2$ &0.089 &0.003  &21.7 &1.1&J23		&RBPL\,J1927$+$6117 &17.7  &0.6$^3$ &0.088 &0.008  &$-$40.6 &2.7  &J22		       \\ 
RBPL\,J1230$+$2518 &15.0  &0.2     &0.055 &0.002  &$-$73.8 &0.9&J23		&RBPL\,J1959$+$6508 &14.4  &0.3  &0.052 &0.005  &$-$25.1 &2.5  &J20		       \\ 
RBPL\,J1238$-$1959 &16.7  &0.2     &0.184 &0.017  &59.9 &2.7&J21		&RBPL\,J2000$-$1748 &17.5  &0.3  &0.130 &0.025  &12.8 &5.3  &J09		       \\ 
RBPL\,J1245$+$5709 &16.9  &0.3$^2$ &0.169 &0.013  &9.7  &2.1&J22		&RBPL\,J2005$+$7752 &15.5  &0.3  &0.047 &0.003  &80.8 &2.2  &J22		       \\ 
RBPL\,J1248$+$5820 &15.0  &0.1$^2$ &0.044 &0.003  &$-$34.8 &2.2&J19		&RBPL\,J2015$-$0137 &16.9  &0.3  &0.149 &0.006  &59.0 &1.3  &J09		       \\ 
RBPL\,J1253$+$5301 &16.4  &0.02$^2$&0.139 &0.003  &39.8 &0.8&J22, J24	&RBPL\,J2016$-$0903 &17.1  &0.3  &0.025 &0.008  &$-$1.0  &8.6  &J10		       \\ 
RBPL\,J1256$-$0547 &15.3  &0.03$^2$&0.232 &0.002  &39.6 &0.4&M30, J23	&RBPL\,J2022$+$7611 &16.0  &0.1  &0.202 &0.004  &35.2 &0.7  &J22, J25     \\ 
RBPL\,J1337$-$1257 &17.7  &0.1     &0.123 &0.026  &80.3 &5.5&J21		&RBPL\,J2030$-$0622 &15.0  &0.5  &0.014 &0.002  &$-$17.7 &3.8  &J10		       \\ 
RBPL\,J1354$-$1041 &16.6  &0.2     &0.004 &0.010  &71.9 &75.8  &J21		&RBPL\,J2030$+$1936 &18.2  &0.4  &0.096 &0.014  &$-$32.3 &4.5  &J19		       \\ 
RBPL\,J1357$+$0128 &17.1  &0.1     &0.137 &0.010  &$-$10.1 &2.1&J24, J26	&RBPL\,J2039$-$1046 &17.4  &0.2  &0.020 &0.008  &44.6 &11.4 &J20		       \\ 
RBPL\,J1419$+$5423 &14.6  &0.01$^2$&0.039 &0.003  &$-$65.8 &1.9&J22		&RBPL\,J2131$-$0915 &17.0  &0.04$^2$&0.082 &0.014  &71.9 &4.7  &J24		       \\ 
RBPL\,J1427$+$2348 &13.7  &0.04$^2$&0.035 &0.001  &$-$54.6 &0.9&M31, J11	&RBPL\,J2143$+$1743 &15.8  &0.04$^2$&0.020 &0.003  &$-$4.3  &4.5  &J10		       \\ 
RBPL\,J1510$-$0543 &17.1  &0.02$^2$&0.014 &0.010  &39.6 &21.8  &J11		&RBPL\,J2146$-$1525 &17.0  &0.1  &0.028 &0.018  &55.5 &19.1 &J24		       \\ 
RBPL\,J1512$-$0905 &15.9  &0.3     &0.028 &0.006  &29.9 &6.2&J11		&RBPL\,J2147$+$0929 &18.4  &0.2$^2$ &0.037 &0.020  &48.8 &16.8 &J19		       \\ 
RBPL\,J1512$+$0203 &16.7  &0.1$^2$ &0.079 &0.007  &$-$29.4 &2.7&J11		&RBPL\,J2148$+$0657 &15.7  &0.02$^2$&0.013 &0.003  &13.7 &5.9  &J10, J23     \\ 
RBPL\,J1516$+$1932 &18.2  &0.1$^2$ &0.012 &0.020  &72.2 &46.1  &J06		&RBPL\,J2149$+$0322 &15.6  &0.1  &0.120 &0.023  &50.6 &11.2 &J11		       \\ 
RBPL\,J1542$+$6129 &14.8  &0.03$^2$&0.092 &0.003  &$-$25.1 &1.0&J19		&RBPL\,J2202$+$4216 &14.0  &0.01$^2$&0.085 &0.001  &$-$11.3 &0.5  &J19, J26     \\ 
RBPL\,J1548$-$2251 &15.8  &0.5     &0.027 &0.011  &$-$69.0 &11.6  &J11		&RBPL\,J2217$+$2421 &17.9  &0.1  &0.183 &0.016  &$-$24.9 &2.5  &J19		       \\ 
RBPL\,J1550$+$0527 &18.1  &0.2     &0.039 &0.021  &$-$42.9 &17.1  &J10, J26	&RBPL\,J2232$+$1143 &16.2  &0.2$^2$ &0.063 &0.005  &$-$7.0  &1.9  &J23		       \\ 
RBPL\,J1555$+$1111 &14.2  &0.1$^2$ &0.030 &0.002  &$-$61.3 &1.7&M31, J10	&RBPL\,J2253$+$1608 &15.2  &0.1  &0.012 &0.007  &$-$39.6 &16.1 &J21		       \\ 
RBPL\,J1558$+$5625 &17.0  &0.5     &0.071 &0.007  &$-$19.7 &2.8&J09		&RBPL\,J2321$+$2732 &18.6  &0.05$^2$&0.043 &0.026  &$-$10.9 &16.4 &J23, J25     \\ 
RBPL\,J1604$+$5714 &17.8  &0.1$^2$ &0.028 &0.010  &$-$26.8 &10.6  &J19		&RBPL\,J2325$+$3957 &17.0  &0.5  &0.036 &0.033  &21.6 &26.5 &J22		       \\ 
RBPL\,J1608$+$1029 &18.1  &0.1$^2$ &0.017 &0.024  &72.2 &39.6  &M31, J11	&RBPL\,J2340$+$8015 &16.6  &0.4  &0.093 &0.008  &$-$36.2 &2.7  &J21		       \\ 
RBPL\,J1635$+$3808 &16.5  &0.01$^2$&0.022 &0.004  &$-$7.7  &5.3&M31, J06	&                   &\mc{2}{c}{} &\mc{2}{c}{}&\mc{2}{c}{}&                         \\
\\\multicolumn{16}{c}{\bf Control Sample Candidates}\\\\[-1.5ex]
J0017$+$8135       &16.4  &0.6     &0.011 &0.005  &72.0 &13.0  &J23           &J1603$+$5730       &17.0  &0.01 &0.018 &0.006  &$-$7.7  &10.0 &J25  \\
J0702$+$8549       &18.3  &0.3     &0.011 &0.013  &$-$31.3 &35.6  &J26           &J1623$+$6624       &18.2  &0.5  &0.032 &0.013  &$-$36.1 &10.9 &J25 \\
J1010$+$8250       &16.2  &0.4     &0.098 &0.011  &$-$19.0 &3.3&J26           &J1624$+$5652       &17.7  &0.1  &0.146 &0.010  &40.8 &1.9  &J25 \\
J1017$+$6116       &16.4  &0.3     &0.015 &0.020  &$-$37.2 &40.0  &J25           &J1638$+$5720       &16.5  &0.2  &0.013 &0.003  &$-$20.7 &7.6  &J26 \\
J1148$+$5924       &13.8  &0.02    &0.025 &0.001  &12.9 &1.4&J25           &J1854$+$7351       &16.3  &0.2  &0.020 &0.004  &$-$22.0 &5.1  &J26 \\
J1436$+$6336       &15.8  &0.6     &0.012 &0.004  &$-$7.5  &10.5  &J25           &J1927$+$7358       &15.6  &0.1  &0.021 &0.005  &$-$30.2 &6.5  &J25 \\
J1526$+$6650       &17.3  &0.2     &0.013 &0.012  &$-$43.6 &27.1  &J27           &J2042$+$7508       &14.3  &0.2  &0.018 &0.001  &$-$9.6  &1.7  &J23 \\
J1551$+$5806       &16.7  &0.04    &0.025 &0.005  &$-$25.5 &5.6&J25           &                   &\mc{2}{c}{} &\mc{2}{c}{}&\mc{2}{c}{}&          \\
\hline                                                                                           
\end{tabular}
\small                                                                                       
\begin{flushleft}                                                                                
$^1$photometry based on USNO-B1.0 R2 unless otherwise noted\\                                         
$^2$photometry based on PTF\\                                                       
$^3$photometry based on USNO-B1.0 R1\\                                          
$^4$Day of June 2013 (JXX) or May 2013 (MXX) in which the observation took place. In cases of multiple measurements we report the dates of the first and the last observations, respectively.                                           
\end{flushleft}                                                                                  
\end{table*}                                                                              

\section{Summary and Discussion}\label{discussion}

We have presented first results from RoboPol, including a linear polarization survey of a
sample of 89 gamma-ray--loud blazars, and a smaller sample of 15 gamma-ray--quiet blazars
defined according to objective selection criteria, easily reproducible in simulations, and
additional unbiased cuts (due to scheduling and quality of observations, independent of
source properties).  These results are therefore representative of the gamma-ray--loud and
gamma-ray--quiet blazar populations, and as such are appropriate for populations studies.

Our findings can be summarised as follows: 
\begin{itemize}
\item The hypothesis that the polarization fractions of gamma-ray--loud and gamma-ray quiet blazars are drawn from the same distribution is rejected at the $3\sigma$ level. 
\item The probability distribution functions of polarization fraction of gamma-ray--loud and gamma-ray--quiet blazars can be well described by exponential distributions. 
\item Using a likelihood analysis we estimate the best-estimate values and $1\sigma$ uncertainties of the mean polarization fraction of each subpopulation, which is the single parameter characterising an exponential distribution. We find $\pmean =6.4
  ^{+0.9}_{-0.8}\times 10^{-2}$ for gamma-ray--loud blazars, and $\pmean =3.2
  ^{+2.0}_{-1.1}\times 10^{-2}$ for gamma-ray--quiet blazars. 
\item The large upwards uncertainty of $\pmean$ for gamma-ray--quiet blazars is a side-effect of the strict cuts we have applied in our likelihood analysis, leaving us only with 7 useable sources for the gamma-ray--quiet sample. This is the reason why the statistical inconsistency between the two populations cannot be also verified with this method. This problem can be improved with a larger gamma-ray--quiet blazar survey, longer integration times, and a more sophisticated analysis. 
\item Polarization angles, $\chi$, for blazars in our survey are consistent with being drawn from a uniform distribution. 
\end{itemize}

It is the first time a statistical difference between the average polarization properties of gamma-ray--quiet and gamma-ray--loud blazars is demonstrated in optical wavelengths. The difference is
consistent with the findings of \citet{Talvikki} for the radio
polarization of gamma-ray--loud and, otherwise similar,
gamma-ray--quiet sources. It thus appears that the gamma-ray--loud
blazars overall exhibit higher degree of polarization in their
synchrotron emission than their gamma-ray--quiet counterparts. One
interpretation for this finding may involve the degree of uniformity
of the magnetic field over the emission region, which is an important
factor affecting the degree of polarization. The bulk of synchrotron
in gamma-ray--loud blazars might therefore originate in regions of
higher magnetic field uniformity than the emission from
gamma-ray--quiet blazars. It is possible that shocks that are strong/persistent enough to accelerate particles capable of gamma-ray emission are also better in locally aligning magnetic field lines and producing regions of high field uniformity, hence a higher polarization degree. 

We have found hints of depolarization at high optical fluxes, an effect that may be attributable to the contribution of unpolarized light to the overall flux by the blazar's host galaxy. The statistics of BL Lac hosts at least are consistent with this idea: in about $50\%$ of the sources studied by \citet{nilsson} the host would have a contribution of more than 50\% the core flux inside our typical aperture. We will examine the effect quantitatively and in more detail using our full first season data in an upcoming publication. 

Inclusion of sources of low (but significantly measured) polarization fraction in the empirical distribution of polarization angles generates some tension (although still not statistically significant) between that distribution and an expected uniform one. This may be a result of foreground polarization at a preferred direction, which, although small and not important for high-polarization sources, tends to align lower polarization sources. Although the sources in the RoboPol sample have been selected to lie away from the Galactic plane so foreground polarization due to interstellar dust absorption should be at a minimum, nearby interstellar material might also induce some degree of foreground polarization. 
For example, such an effect, at the $p \sim 0.8\times10^{-2}$ level, has been seen in the southern sky by \citet{brazilians}. 
A similar level of foreground polarization, $p \sim 0.9\times 10^{-2}$, has been suggested by \citet{fins} for the vicinity of BL Lac (which however lies at relatively low Galactic latitude $b\sim -10^\circ$).  
A cut at $p>3\times10^{-2}$ ensures that sources are intrinsically at least twice as polarized as that, so the effect in measured $\chi$ is minimised.  Because the $13'\times13'$ fields around sources in our monitoring program accumulate exposure during our observing season, we will be eventually able to measure the polarization properties of non-variable, intrinsically unpolarized sources induced by foregrounds to higher accuracy, and better study and correct for this effect in the future. 

For the remainder of the 2013 season we have been monitoring a 3-element sample in linear
polarization with RoboPol: an unbiased gamma-ray--loud blazar sample (51
sources); a smaller, again unbiased, gamma-ray--quiet sample (10 sources);
and a list of high-interest sources that have not made our cuts (24 sources). After the
end of the 2013 season, we will present first light curves and analysis of our sources in
terms of polarization variability and cross-correlations in the amplitude and time
domains. Finally, we will revisit our monitoring sample definition, to strengthen the
robustness of criteria (for example, using RoboPol average R-band fluxes for the
$R$ magnitude cuts), and to develop our automatic scheduling algorithm which aims to
self-trigger high cadence observations during polarization changes that are unusually fast
for a specific source. In this way, we aim to better constrain the linear polarization
properties of the blazar population at optical wavelengths and to provide a definitive
answer to whether a significant fraction of fast polarization rotations do indeed coincide
with gamma-ray flares.

\begin{landscape}
\begin{table}
  \caption{Additional archival information for the 89 high quality sources as well as the
    control sample ones. The table presents only the first 5 lines of the full table which
    is available online as supplementary material. The column numbers
    correspond to individual columns in the online version of the
    table. }
  \label{tab:online}
  \begin{tabular}{llllrlllccllc}
    \hline
1 & 2 & \mcb{3} & \mcb{4} & 5$^{6}$&7 & 8 & 9 $\pm$ 10& 11$_{-12}^{+13}$& 14& 15$\pm$16&17&18\\ 
    ROBOPOL ID     &Survey ID       &\mcb{RA}      &\mcb{DEC}     &\mcb{z}    &\mcb{Class}  &\mcb{Programs}
    &\mcb{$\left<S_{15}\right>_{6-8}$} &\mcb{$m$}        &$F\left(>100\,\rm{MeV}\right)$ &\mcb{$\Gamma$} &VI &In CGRaBS\\  
                   &                &\mcb{(J2000)} &\mcb{(J2000)} &          &            &              &\mcb{(Jy)}                       &\mcb{(fraction)} &($10^{-8}\,\rm{cm}^{-2}\,\rm{s}^{-1}$)   &              &       & \\  
    \hline
    RBPLJ0841$+$7053 &4C\,$+$71.07    &08h41m24.2s  &$+$70d53m41.9s &2.218$^4$   &BZQ         &O, T, F1,2 &2.337$\pm$0.018 &\mcb{\ldots} &6.0 &2.95$\pm$0.07 &91.5           &0 \\ 
    RBPLJ0848$+$6606 &GB6\,J0848+6605 &08h48m54.6s  &$+$66d06m09.6s &\mcb{\ldots} &BZ?         &O          &0.020$\pm$0.002 &\mcb{\ldots} &2.0 &1.96$\pm$0.16 &33.0           &0 \\ 
    RBPLJ0956$+$2515 &OK\,290         &09h56m49.8s  &$+$25d15m15.9s &0.708$^2$   &BZQ         &O          &1.377$\pm$0.022 &0.192$_{-0.009}^{+0.009}$ &2.5 &2.39$\pm$0.07 &84.7 &1 \\
    RBPLJ0957$+$5522 &4C\,$+$55.17    &09h57m38.1s  &$+$55d22m57.0s &0.896$^4$   &BZQ         &O, T       &1.166$\pm$0.007 &0.028$_{-0.003}^{+0.003}$ &8.4 &1.83$\pm$0.03 &23.4 &0 \\
    RBPLJ1014$+$2301 &4C\,$+$23.24    &10h14m46.9s  &$+$23d01m15.9s &0.565$^5$   &BZQ         &O          &1.246$\pm$0.005 &0.085$_{-0.004}^{+0.004}$ &2.4 &2.54$\pm$0.16 &33.0 &1 \\
    \hline                                                                                    
  \end{tabular}
\small
   Column Description: 1: The RoboPol identification name -- 2: A common survey name -- 3, 4: RA, DEC -- 5:
   redshift -- 6: reference for the redshift -- 7: BZCAT Class as of November 14, 2013 with ``BZB'' denoting
   BL Lac objects, ``BZ?'' BL Lac candidates, ``BZQ'' flat-spectrum radio quasars and ``BZU'' blazars of
   uncertain type -- 8: other monitoring programs with ``O'' for OVRO 15 GHz, ``T'' Torun 30 GHz and F1 and F2
   F-GAMMA monitoring before or after June 2009 -- 9: Flux Density at 15 GHz averaged over June-August 2013 --
   10 its uncertainty -- 11: The intrinsic modulation index as computed by Richards et al. 2011 -- 12, 13: the
   lower and upper uncertainty -- 14: Photon Flux above 100\,MeV -- 15, 16: The spectral index and its
   uncertainty as given in the 2FGL -- 17: the
   variability index as given in the 2FGL -- 18: Flag indicating whether the source is in the CGRaBS catalog.\\
   $^{1}$\cite{2013ApJ...764..135S}\\
   $^{2}$\cite{2012ApJ...748...49S}\\
   $^{3}$\cite{2011ApJ...743..171A}\\
   $^{4}$\cite{2010ApJ...715..429A}\\
   $^{5}$\cite{2008ApJS..175...97H}\\
   $^{6}$M. Shaw, personal comm.\\
\end{table}
\end{landscape}

\section*{Acknowledgments}

We thank the referee, Dr. Beverley Wills, for a 
  constructive review that improved this paper. V.P. and E.A. thank
  Dr. F. Mantovani, the internal MPIfR referee, for useful comments on
  the paper. We are grateful to A. Kougentakis, G. Paterakis, and A. Steiakaki,
  the technical team of the Skinakas Observatory, who tirelessly worked
  above and beyond their nominal duties to ensure the timely
  commissioning of RoboPol and the smooth and uninterrupted running of the RoboPol program.  
The U. of Crete group is acknowledging support by the ``RoboPol'' project, which is
implemented under the ``ARISTEIA'' Action of the ``OPERATIONAL PROGRAMME EDUCATION AND
LIFELONG LEARNING'' and is co-funded by the European Social Fund (ESF) and Greek National
Resources. The NCU group is acknowledging support from the Polish National Science Centre
(PNSC), grant number 2011/01/B/ST9/04618. This research is supported in part by NASA
grants NNX11A043G and NSF grant AST-1109911. V.P. is acknowledging support by the European
Commission Seventh Framework Programme (FP7) through the Marie Curie Career Integration
Grant PCIG10-GA-2011-304001 ``JetPop''. K.T. is acknowledging support by FP7 through Marie
Curie Career Integration Grant PCIG-GA-2011-293531 ``SFOnset''. V.P., E.A., I.M., K.T.,
and J.A.Z. would like to acknowledge partial support from the EU FP7 Grant
PIRSES-GA-2012-31578 ``EuroCal''.  I.M. is supported for this research through a stipend
from the International Max Planck Research School (IMPRS) for Astronomy and Astrophysics
at the Universities of Bonn and Cologne. M.B. acknowledges support from the International
Fulbright Science and Technology Award. T.H. was supported in part by the Academy of
Finland project number 267324. The RoboPol collaboration acknowledges observations support
from the Skinakas Observatory, operated jointly by the U. of Crete and the Foundation for
Research and Technology - Hellas.  Support from MPIfR, PNSC, the Caltech Optical
Observatories, and IUCAA for the design and construction of the RoboPol polarimeter is
also acknowledged.

\end{document}